\documentstyle[prl,aps]{revtex}
\begin{document}
\draft

\newtheorem{definition}{Definition}
\newtheorem{theorem}{Theorem}
\newtheorem{lemma}{Lemma}
\newtheorem{conclusion}{Corollary}

\title{Separability of n-particle mixed states:
necessary and sufficient conditions
in terms of linear maps}

\author{Micha\l{}   Horodecki}

\address{Departament of Mathematics and Physics\\
 University of Gda\'nsk, 80--952 Gda\'nsk, Poland}

\author{Pawe\l{} Horodecki}

\address{Faculty of Applied Physics and Mathematics\\
Technical University of Gda\'nsk, 80--952 Gda\'nsk, Poland}

\author{Ryszard Horodecki\footnote{e-mail: fizrh@univ.gda.pl} }

\address{Institute of Theoretical Physics and Astrophysics\\
University of Gda\'nsk, 80--952 Gda\'nsk, Poland}

\maketitle

\begin{abstract}
We provide  necessary and sufficient conditions for separability of mixed
states of n-particle systems. The conditions
are formulated in terms of maps which are positive
on product states of $n-1$ particles.
The method of providing of the
maps on the basis of unextendable product bases
is provided.
The three qubit state problem is reformulated
in the form suggesting possibility of explicite characterisation
of all maps needed for separability condition.
\end{abstract}


\section{Introduction}
Quantum  entanglement \cite{EPR,Schr} is one of the most intriguing
quntum	phenomena leading to various quantum effects with
possible realisations in quantum theory.
The main problem is to find necessary and sufficient conditions
for existence of entanglement of noisy quantum state.
It can be formulated as a problem of serching necessary and
sufficient conditions for separability of the state
or physically speaking - for nonexistence of
quantum correlations in the system state.
The problem is nontrivial even for bipartite
case i. e. the case of quantum system consiting of two subsystems.
The practical, operational solution of the problem
in bipartite case is known
only for the simplest $2 \otimes 2$, $2 \otimes 3$ case
where the positivity of so called partial transposition
has been shown to be necessary and sufficient condition for
separability \cite{Peres,sep}.
For higher dimensional bipartite system it is not the case
\cite{sep,jatran,Choi}.
The general characterisation of separabilty
for those case is given in term of so called positive
maps PM which, however,
have not been operationally characterised so far.
Recently the progress in this direciton has been
made \cite{M1,M2} which suggests that positive maps tests
of separability will have more and more practical applications in
reasonable feature.

In quantum information theory instead of bipartite  case we deal,
in general, with multipartite one like in quantum computing
(see \cite{St98}). Still there is a problem with the noise
as far as practical implementation of quantum computing is concerned.

The definition of multiparticle entanglement
and separability was introduced in \cite{Ve97a,Ve97b}
following the one for bipartite case \cite{Werner}.
The it was considered in case of
distillability of multiparticel entanglement \cite{Mu98,Du99},
separability of very noisy mixed states
\cite{Zy98,Vi99}
and its implications fo NMR quantum computing
\cite{Br99,Sch99,Li99a}. It was also analysed form the point of veiw
of invariants under local fransformations
\cite{Li98,Li99b} and their applications for
quantum cryptography \cite{Ke99}.
The very interesting aspects of multiparticle
mixed states entanglement were studied
 in detailes in context
distillation and of bound entanglement phenomena
\cite{UPB,UPB1,Th99,class},
quantum telecloning \cite{tele},
fragility of entanglement \cite{Ja99}
as well as quantum to classical phase transitions
in quantum computers \cite{Dorit}.

So it is important to develop the formalism which could give general tools
for testing of presence (or absence) of multiparticle entanglement in
given state system.
In this paper we shall make  step in this direciton
firmulationg the necessary and
sufficient condition of separability of
(in general mixed) multiparticle states in terms
of new kind of linear maps
i. e. linear maps positive on product states (LMPP).
We shall also provede the construciton of such maps basing
on UPB \cite{UPB,UPB1} method of generation of bound entanglememt
as well as the technique provieded in \cite{Te98}.
Finally we shall discuss in more detailes the three
qubit case showing that one can hope that some
operational characterisation of LMPP,
(and - equivalently - of the corresponding entanglement witnesess)
is possible.

\section{Repetition: characterisation of bipartite separability - two
dual approaches}
Below we shall briefly racall the way of proving the
mentioned result
as well as the positive maps characterisation of separability
stressing some special aspect of the question.

Let us recall briefly the definition of the
bipartite separability in
finitedimensional case.
\begin{definition}
The state $\varrho$
acting on the Hilbert space ${\cal H}={\cal H}_1 \otimes {\cal H}_2$
is called separable
\footnote{The present definition is due to Werner
\cite{Werner}, who called them classically correlated states.}
if it can be approximated in the trace norm by the states of the form
\begin{equation}
\varrho=\sum_{i=1}^kp_i\varrho_i\otimes\tilde \varrho_i
\label{sep}
\end{equation}
where $\varrho_i$ and $\tilde\varrho_i$ are states on ${\cal H}_1$ and
${\cal H}_2$ respectively.
\end{definition}
Usually we deal with a finite dimensional Hilbert space dim${\cal H}=N<\infty$.
For this case it is known \cite{jatran}
that any separable state can be written as a convex combination of {\it finite}
product pure states $k \leq N^2$, i.e. in those cases the
``approximation'' part of the definition is  redundant.

Now separability can be characterised in two equivalent, though
different ways \cite{sep}.

First we have characterisation in terms of operators
(called, following \cite{Te98},
``entanglement witnesses'')  negative mean values
of which are indicators of entanglement.
We have the fllowing definition (cf. \cite{Te98}):
\begin{definition}
The observable $W$ is called entanglement witness iff
its mean values on all separable states are nonnegative i. e.
iff ${\rm Tr}(\varrho W)\geq 0$ for all separabile $\varrho$.
\end{definition}

Now we have the following characterisation of separability \cite{sep}:

{\bf Characterisation (I)} $\varrho$ is separable iff
${\rm Tr}(\varrho W)\geq 0$ for all entanglement witnesses $W$.

On the other hand we have the characterisation in terms of positive maps
which are defined as follows (see \cite{Wo74,Ja72}):
\begin{definition}
Let ${\cal B}({\cal H})$ stand for the algebra of all
operators on Hilbert space ${\cal H}$, ${\rm dim}{\cal H} <\infty$.
The the map $\Lambda:{\cal B}({\cal H}_2) \rightarrow {\cal B}({\cal H}_1)$
is positive  iff $\Lambda(A)\geq 0$ for all $A \geq 0$,
$A \in {\cal B}({\cal H}_1)$.
\end{definition}

Recall that $X \geq 0$ stands for hermitian operator with nonegative
eigenvalues and such operator is called {\it positive}.

Now we can recall the second {\it characterisation
of separability in terms of positive maps}

{\bf Characterisation (II)} \cite{sep} $\varrho$ defined on
${\cal H}_{1} \otimes {\cal H}_{2}$ is separable iff
operator $I \otimes \Lambda(\varrho)\geq 0$ for all positive $\Lambda$.

{\bf Remark .-}
The above characterisation relies on the fact that there are the
positive maps $\Lambda$ such that their multiplication by identity $I \otimes
\Lambda$ is not positive. The well known map ot that kind
is simply transposition $T$ \cite{Ja72,Wo74} defined
on martix elements as $[ T(A) ]_{mn}=[ A ]_{nm}$.
The map $I \otimes T$ is not positive and the condition
$I \otimes T (\varrho) \geq 0$ is the form of separability
criterion introduced by Peres \cite{Peres}.

What is the difference betweeen the characteriastions
(I) and (II) ? The first has {\it a scalar}
character as it utilises conditions
imposed on mean values, while the second
one applies the {\it operator} conditions imposing
the boundaries on the spectrum of some operators.

Being both characterisation of the same set the
conditions (I) and (II) are both equivalent
if considered collectively.
The link is fomed by the fact that there is one to one isomorphism
between positive maps and entanglement witnesses $A$ coming form
Ref. \cite{Ja72}. Namely
\begin{equation}
A \leftrightarrow I \otimes \Lambda_{A}(P_{+})
\label{isom}
\end{equation}
Where $P_{+}=| \Psi\rangle\langle \Psi |$,
$|\Psi\rangle=\sum_{i=1}^{dim {\cal H}_{1}} |k \rangle \otimes |k \rangle$,
$|\Psi\rangle \in {\cal H}_{1} \otimes {\cal H}_{1}$.
What is the motivation to look rather positive maps then
entanglement witnesses ?
Though characterisations (I) and (II) are completely equivalent
the single conditions (i) $Tr(\varrho A)\geq 0$
(ii) $I \otimes \Lambda_{A}(A) \geq 0$ are not.
In particular sometimes the second one is much stronger than the first one.
Let us recall the $2 \otimes 2$ example corresponding to the system
composed of two $\frac{1}{2}$-spin systems.

{\it Example.} Let $A=V$ where $V$ is a ``flip'' operator i. e.
$V |\phi \rangle \otimes |\psi \rangle= |\psi \rangle \otimes |\phi\rangle $.
It has been shown \cite{Werner} that $V$ is a legitimite entanglement witness
i. e. it gives nonegative mean value on product states, ergo
has the same property on all separable states.
On the other hand it detects entanglement of
all states $\sigma$ with property
$\langle \Psi_{-}| \sigma |\Psi_{-} \rangle>\frac{1}{2}$,
$|\Psi_{-} \rangle =
\frac{1}{\sqrt{2}}(|01\rangle - |10\rangle)$
Hence $Tr(A \varrho)\geq 0$ is nontrivial separability condition.
However it is weak condition - it {\it does not} detect
for example entanglement of {\it maximally entangled} state
$\frac{1}{\sqrt{2}}(|00\rangle + |11\rangle)$
giving on it strictly positive value $1$.

But let us apply isomorphism (\ref{isom}).
Then we get the map $V \rightarrow \Lambda_{V}=T$
which is the transposition.
But it was shown \cite{sep} that $I \otimes T (\varrho) = I \otimes \Lambda_{V}(\varrho)
\geq 0$ is {\it sufficient } condition
for separability of mixed states hence it detect {\it all}
entanglement.
Thus in this case single condition from
Characteristion (II) appeared to be
much more powerfull than form (I).
In this way we have seen the main motivation to study the problem
maps dececting entanglement in place of entanglement witnesses.

Finally we recall that there are two other weaker notions introduced in context of
separability. Namely we have the definition (see \cite{UPB}):
\begin{definition}
The state of n-partite system is called semiseparable
iff for any index $m$ ($1\leq m \leq n$)
the state can be represented as a separable state
of bipartite system composed form $m$-th subsystem
and the rest of the system:
\begin{equation}
\varrho=\sum_{i=1}^{N}p_{i}\varrho^{i}_{m} \otimes \varrho^{i}_{1, ..., m-1, m+1, ...n}.
\end{equation}
\end{definition}
It has been shown that there are semiseparable states
whic are not separable \cite{UPB,UPB1}.
The characterisations I, II
recalled above can be easily applied
to give necessary and sufficient conditions of semiseparability.
It is enough to divide system into proper subsystems
and apply the condition, then divide into two other ones and
apply the corresponding separability condition again, etc.

Finally there is another definition
concerning very interesting property.
Namely one can imagine that we have entanglement in the system
composed form $n$ elementary subsystems but
it is represented by ``clusters'' of some restricted size
(in number of elementary subsystems involved).
We have the corresponding definition \cite{Ve97b}:
\begin{definition}
State $\varrho$ does not represent entanglement
of not more than $k$ elementary subsystems iff
it admits decomposition
\begin{equation}
\varrho=\sum_{i=1}^{N}p_{i}\varrho^{i,1} \otimes ... \otimes
\varrho^{i,n_{i}}
\end{equation}
and $\varrho^{i,n_{i}}$ describes subset of not more than
$k$ subsystems form all $n$ elementary ones.
\end{definition}
We discuss the above notion in the context of the present results.

\section{Necessary and sufficient conditions
 for separability of n-particle systems}
The definition of separable states \cite{Werner} was easily generalized
to systems composed of more than two subsystems (see for example
\cite{volume}):
\begin{definition}
The state $\varrho$
acting on the Hilbert space ${\cal H}=\mathop{\otimes}
\limits_{l=1}^{n} {\cal H}_l$
is called separable
if it can be approximated in the trace norm by the states of the form
\begin{equation}
\varrho=\sum_{i=1}^k p_i \mathop{\otimes}\limits_{l=1}^{n}
\varrho^{l}_i
\label{sep2}
\end{equation}
where $\varrho^{l}_i$ are states on ${\cal H}_l$ .
\end{definition}
Straightforward generalization of the proof about decomposition
from Ref. \cite{jatran} gives us the possibility of omitting the
approximation part in the definition:
\begin{lemma}
Any separable state $\varrho$ of a system composed by $m$
subsystems can be written as:
\begin{equation}
\varrho=\sum_{i=1}^k p_i P^{i}_{prod}, \  k \leq N^2
\end{equation}
where $P^{i}_{prod}$ are  pure product states
having	the n-decomposable form
$\mathop{\otimes}\limits_{l=1}^{m} P_l$, where $P_l$ are projectors acting on ${\cal H}_l$.
\end{lemma}

Now following the proof form
the Ref. \cite{sep} can provide the simple generalisation of
the Characterisation (I) form the previous section.

\begin{theorem}
State $\varrho$ defined on
${\cal H}={\cal H}_{1} \otimes {\cal H}_{2} \otimes ... {\cal H}_{n}$
is separable iff
$Tr(A \varrho)\geq 0$ holds for all observables $A$ on $\cal H$
such that
\begin{equation}
Tr (A P_{1} \otimes P_{2} \otimes ... \otimes P_{n}) \geq 0
\label{witn}
\end{equation}
for all product projections $P_{1} \otimes P_{2} \otimes ... \otimes P_{n}$
($P_{i}$ are projectors on ${\cal H}_{i}$).
\label{ch1}
\end{theorem}

{\it Remark.-}
As the conclusion from the Hahn-Banach theorem
is valid for {\it any} Banach space our
theorem can be generalized  for infinitely dimensional Hilbert spaces.

Now we provide the following theorem which is nontrivial generalisation
of Characterisation (II) to the multiparticle case.
\begin{theorem}
State $\varrho$ defined on
${\cal H}= {\cal H}_{1} \otimes {\cal H}_{2} \otimes ... {\cal H}_{n}$
is separable iff
$I \otimes \tilde{\Lambda} (\varrho)\geq 0$
for all LMPP $\tilde{\Lambda}$ i. e. all linear maps
$\tilde{\Lambda}: {\cal B}({\cal H}_{2} \otimes ... {\cal H}_{n})
\rightarrow {\cal B}({\cal H}_{1})$
such that
\begin{equation}
\tilde{\Lambda}(P_{2} \otimes ... \otimes P_{n})\geq 0
\label{mapdod}
\end{equation}
for all product projections
$P_{2} \otimes ... \otimes P_{n}$
($P_{i}$ are projectors on ${\cal H}_{i}$).
\label{thmaps}
\end{theorem}

{\it Proof .-}
We shall prove only the ``if'' part as the only if is trivial.
We shall show that iff the condition (\ref{mapdod})
holds for all required $\tilde \Lambda$ than
(\ref{witn}) takes place for all entanglement witnesses $A$.
Let (\ref{mapdod}) hold for all maps $\tilde \Lambda$.
Than in particular $Tr (P_{+} \tilde{\Lambda}(\varrho)) \geq 0$
or $Tr (I \otimes (\tilde\Lambda)^{\dagger} (P_{+}) \varrho )\geq 0$.
Hence the assumption (\ref{mapdod}) leads to the condition
$Tr(\tilde{A} \varrho) \geq 0$ for
\begin{equation}
\tilde{A}=I \otimes (\tilde\Lambda)^{\dagger} (P_{+}).
\label{forma}
\end{equation}
Now it is enough to show that any entanglement witness
is of the above form for some $\Lambda^{\dagger}$ such that
$I \otimes \Lambda^{\dagger}(P_{2} \otimes ... \otimes P_{n}) \geq 0$
for all $P_{i}$ being projections on ${\cal H}_{i}$.
Take any entanglement witness $A$.
Then it must satisfy (\ref{witn}) and on the other hand it must be of the form
(\ref{isom}) for some map $\Lambda:
{\cal B}({\cal H}_{1}) \rightarrow
{\cal B}({\cal H}_{2} \otimes ... \otimes {\cal H}_{n})$.
\footnote{The latter comes form the fact that (\ref{isom})
gives general isomorphism between linear maps and hermitian operators.}
Let design by $\Lambda_{1}=\Lambda^{\dagger}$.
Then the condition (\ref{witn}) is equivalent
to the form
\begin{equation}
Tr(I \otimes \Lambda_{1}^{\dagger} (P_{+})
P_{1} \otimes P_{2} \otimes ...\otimes P_{n})=
Tr (P_{+} P_{1} \otimes \Lambda_{1}(P_{2} \otimes ...\otimes P_{n}))\geq 0
\end{equation}

Now let us recall that (i) $P_{+}^{T_{1}}=V$ where $V$ is a flip
operator acting on ${\cal H}_{1} \otimes {\cal H}_{1}$
\footnote{Here $X^{T_{1}} \equiv T \otimes I (X)$
i. e. partial transposition with respect to the first Hilbert
space ${\cal H}_{1}$.}
(ii) $Tr(XY)=Tr(X^{T_{1}} Y^{T_{1}})$,
(iii) \cite{Werner} $Tr(V X \otimes Y)= Tr (X^{T_{1}}Y)$.
Application of those properties leads us to the conclusion that the above
equation is equivalent to the one
\begin{equation}
Tr(P_{1} \Lambda_{1}(P_{2} \otimes ... \otimes P_{n})) \geq 0
\end{equation}
Form arbitrariness of projectors $P_{i}$ we get
immediately that  $\Lambda_{1}$ posess the property of
$\tilde\Lambda$ maps form the theorem (\ref{thmaps}).
Hence any entanglement witness is of the form \ref{forma}
which ends the proof.

\section{Multiparticle case - construction based on unextendible
product bases}
In this section we shall utilise a pioneering method of Ref.
\cite{Te98} to produce  maps positive on product states of
two q-bits. The method was the following producing the
positive maps which are indecomposable.
Namely it has been shown \cite{UPB,UPB1} that there exist so called unextendable product
bases (UPB) i. e.  such sets of product vectors
in ${\cal H}={\cal H}_{1} \otimes {\cal H}_{2} \otimes ... \otimes {\cal H}_{n}$
that (i) there os no product vector orthogonal to them
(ii) they do not span the whole space ${\cal H}$.
It produced systematic way of generating so called
bound entangled states i. e.  entangled states form which no
entanglement can be distilled.
The systematis method is as follows: find UPB set
then take the projector $P_{UPB}^{\perp}$ onto space orthogonal to the set
and normalise it. The resulting state $\varrho_{UPB}$
is bound entangled. In particular it has PPT property with respect to any
division of the system onto two parts.
For bipartite $\varrho_{UPB}$ their structure was utilised
to get indecomposable maps in the following way \cite{Te98}:
take the projector onto the UPB set $P_{UPB}$ of
bipartite states, calculate minimal value of
$\epsilon=\mathop{max}\limits_{P\otimes Q} Tr(P_{UPB} P \otimes Q)$.
Then take maximally entangled state $\Psi_{+}$ and
calculate maximal value $c=\mathop{max}\limits_{P \otimes Q} Tr(P_{UPB} P \otimes Q)$
which is strictly positive. Finally the following operator
\begin{equation}
A=P_{UPB}-\frac{\epsilon}{c}P_{+}
\end{equation}
is an entanglement witness for $\varrho_{UPB}$ i. e. it is
has positive mean value on all product states but negative
on $\varrho_{UPB}$. The corresponding linear map
generated automatically via isomorphism (\ref{isom})
is positive indecomposable map (see \cite{Te98} for details).

Below we shall show that the same way can lead to
production of positive maps on products
if applied to multipartite systems
\footnote{The first part of reasoning presented below (concerning
entanglement witnesses) was independently performed
by B. Terhal \cite{private}}.
Consider any multipartite system and its bound entangled state
$\varrho_{UPB}$ produced by given UPB set.
Then take the following operator
\begin{equation}
\tilde{A}\equiv P_{S_{UPB}}-\frac{\epsilon}{c}C
\end{equation}
where $C\geq 0$ is an arbitrary positive operator
with $Tr(P_{S_{UPB}}^{\perp} C) \geq 0$
and
$c=\mathop{max}\limits_{P_1 \otimes ... \otimes P_{N}}
Tr(C P_1 \otimes ... \otimes  P_N)$. Then the operator
$\tilde{A} \in {\cal B}({\cal H}_{1}\otimes ... \otimes H_{N})$
has by the very definition positive mean values on product states.
So it satisfies conditions of the Theorem \ref{thmaps}.
On the other hand it has negative mean value on
$\varrho_{UPB}$:
\begin{equation}
Tr(\tilde{A} \varrho_{UPB})=-\frac{\epsilon}{c}< 0
\end{equation}
being then entanglement witness for this state.
By the theorem that the corresponding linear
map $\Lambda_{\tilde{A}}$ (this time {\it via} relation (\ref{forma})
is postive on product states. On the other hand
it detects multipartite entanglement
as $\langle \Psi_{+}| I_1 \otimes \Lambda_{\tilde{A}}
\varrho_{UPB} |\Psi_{+}\rangle= Tr(\tilde{A} \varrho_{UPB})< 0$.
The above construction if applied of
the states generated by the most simple three qubits UPB
\cite{UPB} $S_{UPB}=span \{ |000\rangle, |+1-\rangle, |1-+\rangle, |+-1\rangle \}$.
(with $|\pm\rangle=\frac{1}{\sqrt{2}}(|0\rangle\pm |1\rangle)$)
leads to the map which is, however,  of complicated form.
Below we shall analyse in more detailes
the properties of maps revealing entanglement of three qubits.

\section{Three q-bit case - general analysis}
Consider now again the case of three q-bits i. e.
the system defined on a Hilbert space ${\cal H}={\cal H}_{1} \otimes
{\cal H}_{2}\otimes {\cal H}_{2}={\cal C}^{2}\otimes  {\cal C}^{2}\otimes
{\cal C}^{2}$.
We have the following obesrvation:

{\bf Observation .-} Any entanglement witness
(\ref{mapdod})
$\tilde{\Lambda}: {\cal B}({\cal C}^{2} \otimes {\cal C}^{2})
\rightarrow {\cal B}({\cal C}^{2})$ satisfies
$\forall$ $\sigma_{a}, \sigma_{b}, \sigma_{c}$ ($\sigma_{x}$ -
one q-bit states)
\begin{equation}
Tr[\sigma_{a}\tilde{\Lambda}(\sigma_{b} \otimes \sigma_{c})] \geq 0
\label{o1}
\end{equation}
or, equivalently,
\begin{equation}
Tr[\tilde{\Lambda}^{\dagger}(\sigma_{a})\sigma_{b} \otimes \sigma_{c}] \geq 0
\label{o2}
\end{equation}

Now we have the following theorems:
\begin{theorem}
The three q-bit state $\varrho$ is separable
iff $I \otimes \tilde{\Lambda}(\varrho) \geq 0$
for all
$\tilde{\Lambda}: {\cal B}({\cal C}^{2} \otimes {\cal C}^{2})
\rightarrow {\cal B}({\cal C}^{2})$
 satisfies  $\forall \sigma$ ($\sigma$ one qubit state)
 \begin{equation}
\tilde{\Lambda}(\sigma)=A+B^{T_{2}}
\label{AB}
 \end{equation}
 for some positive $A$ and $B$.
\end{theorem}

Below we shall write down the problem solution of which would
be sufficient for characterisation of tripartite entanglement.
Namely {\it all} the maps we need are the maps acting in the following
way
\begin{equation}
\tilde{\Lambda}[(I \otimes \hat{n}\vec{\sigma})
\otimes (I \otimes \hat{m}\vec{\sigma})]=\alpha(
\hat{n},\hat{m})[I + \vec{k}(\hat{n},\hat{m})\vec{\sigma}]
\end{equation}

where $\hat{x}\vec{\sigma}\equiv \sum_{i=1}^{3},
x_{i}\sigma_{i}$ where $\{ x_{i} \}$ represent coordinates of real unit
vector $\hat{x}$, while $\sigma_{i}$ stands for Pauli matrices.
The vector $\vec{k}$ must satisfy $||\vec{k}||\leq 1$ and
$\alpha \geq 0$. It is not easy to see whether it
is possible to characterise
the maps of that form. Nevertheless this is the first time
when aiming in characterisation of the map with respect
to somehow positivity
we have clear positivity conditions - {\it via}
treedimensional Bloch vectors) - on {\it both its arguments and the results}.
It follows from the fact that triparticle separability condition
for $2 \otimes 2 \otimes 2$ system is {\it more} restrictive
then separability of $2 \otimes 4$ system.
Actually there is {\it more} maps of that kind.
This can be simply argued from the fact that three q-bits semiseparability
(i.e. separability of state on ${\cal C}^{2} \otimes {\cal C}^{2}
\otimes {\cal C}^{2}$ with respect to all ``cuts''
into two parts of type ${\cal C}^{2} \otimes {\cal C}^{4}$)
is not equivalent to separability \cite{UPB}
( we shall use this fact in the next section).
As there is more maps positive on product states
then the positive ones
it may happen that the set of LMPP, strictly larger than the
set of PM, is also more regular and possible to characterise
analytically.
\section{Discussion}
We have provided the necessary and sufficient conditions
of separability of multiparticle quantum systems
in terms of maps positive on product states.
This is a new class of maps forming a subset of set
of positive maps. The construciton of examples basing on existence
of unextendable product bases as well as the corresponding bound
entangled states has been provided. For the case of three q-bits
the necessary and sufficient condition for maps of the form can be
formulated using the Bloch vector formalism which allow to hope that
some operational charaterisation of the of the set of the maps is possible.

Let us outline the problem for further research
in the context of the present results.
As we have mentioned, the previous results can be easily applied to
characterise semiseparability.
The present ones can be applied to verions of partial semiseparability
i. e. when systems is divided into, for example, three parts
and is to be separable with respect to such partition).
there is however an open problem how to apply it for
the third definition (involving ``clusters'' of size $k$,
see section II). In fact, the naive application of the
present criteria should, in principle fail.
Indeed, the simple inspection shows that it can happen
that clusters can be shifted. Due to this fact,
if applying the present approach naively,
one can have fake detection of entanglement
of, say, $k+1$ subsystems while only $k$-size entanglement is present.
So there is need for further investigation towards characterisation of
separability in sense of definition 3.

In the present paper we have an example of case when physical
problem - characterisation of multiparticle separability
(or - equivalently - entanglement) leads to new mathematical objects
linear maps positive on products -
investigation structure of which can be interesting
itself. Moreover we know that, in general,
single maps condition is much more powerfull that
the entanglement witness condition
so after finding entanglement witness for multiparticle case
explicite generation of linear maps positive on product states (LMPP)
may give much stronger separability conditions too
which is of practical importance.

\end{document}